\documentclass[12pt]{article}

\catcode`\@=11
\@addtoreset{equation}{section}

\global\arraycolsep=2pt

\usepackage{mathrsfs,amsbsy,amssymb,latexsym,amsfonts,amsmath,cite}
\usepackage{graphicx,color}

\allowdisplaybreaks

\newcommand{\diag}{{\rm{diag}} }

\newcommand{\al}{\alpha}
\newcommand{\be}{\beta}
\newcommand{\ga}{\gamma}
\newcommand{\de}{\delta}
\newcommand{\ep}{\epsilon}

\newcommand{\str}{{\rm STr}}

\newcommand{\alg}[1]{\mathfrak{#1}}
\newcommand{\el}{\nonumber}
\newcommand{\nln}{\nonumber\\}

\begin{document}

\begin{flushright}
\parbox{4cm}
{KUNS-2506}
\end{flushright}

\vspace*{1.5cm}

\begin{center}
{\Large\bf 
Yang-Baxter deformations and string dualities} 
\vspace*{1.5cm}\\
{\large Takuya Matsumoto$^{\dagger}$\footnote{
E-mail:~takuya.matsumoto@math.nagoya-u.ac.jp} 
and Kentaroh Yoshida$^{\ast}$\footnote{
E-mail:~kyoshida@gauge.scphys.kyoto-u.ac.jp}} 
\end{center}
\vspace*{0.25cm}
\begin{center}
$^{\dagger}${\it 
Institute for Advanced Research and  
Department of Mathematics, \\ 
Nagoya University, \\ Nagoya 464-8602.Japan.} 
\vspace*{0.25cm}\\ 
$^{\ast}${\it Department of Physics, Kyoto University, \\ 
Kyoto 606-8502, Japan.} 
\end{center}
\vspace{1cm}

\begin{abstract}
We further study integrable deformations of the AdS$_5\times$S$^5$ superstring 
by following the Yang-Baxter sigma model approach with classical $r$-matrices satisfying 
the classical Yang-Baxter equation (CYBE). Deformed string backgrounds specified 
by $r$-matrices are considered as solutions of type IIB supergravity, and therefore  
the relation between gravitational solutions and $r$-matrices may be called 
the gravity/CYBE correspondence. 
In this paper, we present a family of string backgrounds associated with  
a classical $r$-matrices carrying two parameters and its three-parameter generalization. 
The two-parameter case leads to the metric and NS-NS two-form of a solution found by 
Hubeny-Rangamani-Ross [hep-th/0504034] and another solution in [arXiv:1402.6147]. 
For all of the backgrounds associated with the three-parameter case, 
the metric and NS-NS two-form are reproduced by performing TsT transformations 
and S-dualities for the undeformed AdS$_5\times$S$^5$ background. 
As a result, one can anticipate the R-R sector that should be reproduced via 
a supercoset construction. 
\end{abstract}

\setcounter{footnote}{0}
\setcounter{page}{0}
\thispagestyle{empty}

\newpage

\tableofcontents

\section{Introduction}

The AdS/CFT correspondence \cite{M} has a significant property, integrability \cite{review}. 
It enables us to compute exactly some quantities such as scaling dimensions 
of composite operators and scattering amplitudes at arbitrary couplings   
even though those are not protected by supersymmetries. On the string-theory side, the integrability 
has an intimate relation to classical integrability of non-linear sigma models in two dimensions. 
In particular, the mathematical structure of the target-space geometry is closely associated with the integrability. 

\medskip 

The supergeometry of AdS$_5\times$S$^5$ is represented by the following supercoset, 
\begin{eqnarray}
\frac{PSU(2,2|4)}{SO(1,4)\times SO(5)}\,. \label{coset}
\end{eqnarray} 
The Green-Schwarz type action is constructed based on this supercoset \cite{MT}.  
It is known that the coset (\ref{coset}) enjoys the $\mathbb{Z}_4$-grading property 
and it ensures the classical integrability \cite{BPR}. For another formulation \cite{RS}, 
the classical integrability is discussed in \cite{Hatsuda}. 

\medskip 

What kinds of string backgrounds would be concerned with integrability?  
It is well-known that symmetric cosets in general lead to classical integrability. 
For the symmetric cosets, consistent string backgrounds have been  
classified in \cite{Zarembo-symmetric,Wulf}. 
On the other hand, some of non-symmetric cosets may lead to integrable sigma models, 
but there is no general prescription. 
Thus one has to study case by case and hence the main focus has been on simple cases. 
In the case of squashed S$^3$\,, a $q$-deformation of $\mathfrak{su}(2)$ and its affine extension 
have been shown in \cite{KY,KYhybrid,KMY-QAA}. Similar results have also been obtained 
for 3D Schr\"odinger spacetimes \cite{KY-Sch,ORU}. 
For an earlier attempt to study higher-dimensional cases, see \cite{BR}. 
For integrable deformations of Wess-Zumino-Novikov-Witten models, see \cite{KOY,DMV4,Hollowood,Sfetsos}. 

\medskip 

A systematic method to study integrable deformations was proposed by Klimcik \cite{Klimcik}. 
The original form was applicable only to principal chiral models 
and it was based on classical $r$-matrices satisfying modified classical Yang-Baxter equation (mCYBE). 
Then it was generalized to symmetric coset cases \cite{DMV}. Just after that, 
it was further applied to a $q$-deformation of the AdS$_5\times$S$^5$ superstring \cite{DMV2}. 
The $q$-deformed metric (in the string frame) and the NS-NS two-form were derived in \cite{ABF}. 
Notably, the metric exhibits a singularity surface. 
A generalization to other cases are studied in \cite{HRT}. A mirror description has been proposed 
in \cite{AdLvT,AvT}. A possible resolution of the singularity has been argued 
by taking the fast-moving string limit \cite{Kame}. 
Giant magnon solutions are studied in \cite{AdLvT,Magnon}. 
Deformed Neumann models are constructed \cite{deformed-Neumann}. 
A new coordinate system has been argued in \cite{Kame2}. 
Although it seems quite difficult to fix the dilaton and the R-R sector for the $q$-deformed 
AdS$_5\times$S$^5$ superstring, the dilaton was determined at least in lower-dimensional cases 
such as AdS$_2\times$S$^2$ \cite{LRT}. 
For two-parameter generalizations, see \cite{HRT,Hoare}. 

\medskip 

It is also possible to consider another kind of integrable deformations of the AdS$_5\times$S$^5$ superstring 
based on the classical Yang-Baxter equation (CYBE) \cite{KMY-Jordanian-typeIIB}, rather than mCYBE. 
We have already found out some classical $r$-matrices, which correspond to solutions of type IIB supergravity. 
An example is the Lunin-Maldacena-Frolov backgrounds \cite{LM,Frolov} 
and the corresponding $r$-matrix is composed of the Cartan generators of $\alg{su}(4)$ \cite{LM-MY}. 
Another example is gravity duals for non-commutative gauge theories \cite{HI,MR,DHH,MS}. 
The associated $r$-matrices are of peculiar Jordanian type \cite{MR-MY}. A rather 
remarkable example is the AdS$_5\times T^{1,1}$ case. The $T^{1,1}$ background 
is argued to be non-integrable \cite{BZ}. On the other hand, the $T^{1,1}$ geometry can be represented by a coset 
and hence one may consider Yang-Baxter deformations of $T^{1,1}$\,. Then it has been shown in \cite{CMY} 
that the resulting geometry nicely agrees with $\gamma$-deformations of $T^{1,1}$ previously discussed in \cite{LM,CO}. 
For a short summary of the works based on the CYBE, see \cite{MY-summary}. 

\medskip 

In this paper, we further study integrable deformations of the AdS$_5\times$S$^5$ superstring 
along the line of \cite{KMY-Jordanian-typeIIB}. Deformed string backgrounds specified by $r$-matrices 
are considered as solutions of type IIB supergravity, and therefore the relation between 
gravitational solutions and $r$-matrices may be called the gravity/CYBE correspondence. 
We will present here a family of string backgrounds associated with  
a classical $r$-matrices carrying two parameters and its three-parameter generalization. 
The two-parameter case leads to the metric and NS-NS B-field of a solution found by 
Hubeny-Rangamani-Ross \cite{HRR} and another solution in \cite{SUGRA-KMY} as special cases. 
More generally, for all of the backgrounds associated with the three-parameter case, 
the metric and NS-NS two-form are reproduced by performing TsT transformations 
and S-dualities for the undeformed AdS$_5\times$S$^5$ background. 
As a result, one can anticipate the R-R sector that should be reproduced via 
a supercoset construction.

\medskip 

A remarkable point is that the solution obtained in \cite{SUGRA-KMY} has been reproduced 
by performing a duality chain for the AdS$_5\times$S$^5$, and as a result it is shown to be 
a consistent string background. Namely, it is automatically ensured that the beta function vanishes.  
This point should be stressed against those who are skeptical about the solution in \cite{SUGRA-KMY}. 

\medskip 

This paper is organized as follows. Section \ref{sec:YBsigma} provides a short review of 
Yang-Baxter deformations of the AdS$_5\times$S$^5$ superstring. We present a classical $r$-matrix 
with two parameters, which is a solution of the CYBE. 
Then the deformed metric and NS-NS two-form are obtained by evaluating the classical action. 
In section \ref{sec:chain}\,,  we reproduce the resulting metric and NS-NS two-form 
by performing a chain of dualities for the AdS$_5\times$S$^5$ background. 
Section \ref{sec:summary} generalizes the previous argument  
to a three-parameter case and considers the correspondence between 
the deformations of AdS$_5\times$S$^5$ 
and the associated classical $r$-matrices. 
Section \ref{sec:concl} is devoted to conclusion and discussion.

\medskip 
 
In Appendix \ref{app:notation}, our notation and convention are summarized. 
The Buscher rules of T-duality are listed in Appendix \ref{app:Tdual}.

\section{Jordanian deformations of the AdS$_5\times$S$^5$ superstring} 
\label{sec:YBsigma}

In subsection \ref{subsec:YBsigma}\,, let us first recall the formulation of Yang-Baxter sigma models. 
Then we consider a two-parameter deformation of the AdS$_5$ in subsection
\ref{subsec:2-paraAdS}\,.
The resulting metric and NS-NS two-form are computed in subsection
\ref{subsec:metric}\,. 

\subsection{Yang-Baxter deformations of the AdS$_5\times$S$^5$ superstring} 
\label{subsec:YBsigma}

The Yang-Baxter sigma model approach \cite{Klimcik,DMV,DMV2} is applicable to 
the AdS$_5\times$S$^5$ superstring by using a classical $r$-matrix satisfying the CYBE 
\cite{KMY-Jordanian-typeIIB}. 
Then the deformed action is given by 
\begin{align}
S=-\frac{1}{4}(\ga^{\al\be}-\ep^{\al\be}) 
\int^\infty_{-\infty}\!\! d\tau \int^{2\pi}_{0}\!\! d\sigma\,   
\str\Bigl(A_\al d\circ \frac{1}{1-\eta R_g\circ d}A_\be \Bigr)\,, 
\label{action}
\end{align}
where the left-invariant one-form is defined as  
\begin{align}
A_\al\equiv g^{-1} \partial_\al g\,, \qquad  g\in SU(2,2|4)\,.  
\end{align}
By taking the parameter $\eta\to 0$, 
the action \eqref{action} reduces to the undeformed one\cite{MT}. 
Here the flat metric $\ga^{\al\be}$ and the anti-symmetric tensor 
$\ep^{\al\be}$ on the string world-sheet
are normalized as $\ga^{\al\be}={\rm diag}(-1,1)$
and $\ep^{\tau\sigma}=1$\,, respectively.  
The operator $R_g$ is defined as 
\begin{align}
R_g(X)\equiv g^{-1}R(gXg^{-1})g\,, 
\end{align}
where a linear operator $R$ is a solution of CYBE rather than mCYBE.  
The R-operator is related to a classical $r$-matrix in the tensorial notation through 
\begin{align}
&R(X)=\str_2[r(1\otimes X)]=\sum_i \bigl(a_i\str(b_iX)-b_i\str(a_iX)\bigr) 
\label{linearR} \\
&\text{with}\quad r=\sum_i a_i\wedge b_i\equiv \sum_i (a_i\otimes b_i-b_i\otimes a_i)\,. \el 
\end{align}
The generators $a_i, b_i$ are some elements of $\alg{su}(2,2|4)$\,. 
The operator $d$ is defined by
\begin{align}
d\equiv P_1+2P_2-P_3\,,  
\end{align}  
with projectors $P_k$ ($k=0,1,2,3$) from $\alg{su}(2,2|4)$
to its $\mathbb{Z}_4$-graded components $\alg{su}(2,2|4)^{(k)}$\,. 
In particular, $\alg{su}(2,2|4)^{(0)}$ is a gauge symmetry, $\alg{so}(1,4)\oplus\alg{so}(5)$\,.

\subsection{A two-parameter deformation of AdS$_5$ }
\label{subsec:2-paraAdS}

We consider here a deformation of the AdS$_5$ bosonic part of \eqref{action} 
based on the following classical $r$-matrix of Jordanian type,  
\begin{align}
r_{\rm Jor}=E_{24}\wedge (c_1 E_{22} - c_2E_{44})\,, 
\label{Jor-r}
\end{align}
where $E_{ij}$ ($i,j=1,2,3,4$) are the fundamental representation of the generators 
of $\alg{su}(2,2)$ defined by $(E_{ij})_{kl}=\de_{ik}\de_{jl}$\,. Here $c_1$ and $c_2$ 
are constant complex numbers.

\medskip 

To evaluate the action, it is convenient to rewrite the metric part and NS-NS two-form coupled part 
of the Lagrangian \eqref{action} into the following form,  
\begin{align}
L_G&=\frac{1}{2}\left[A_\tau P_2(J_\tau)-A_\sigma P_2(J_\sigma)\right]\,, \el \\
L_B&=\frac{1}{2}\left[A_\tau P_2(J_\sigma)-A_\sigma P_2(J_\tau)\right]\,, 
\label{LGLB} 
\end{align} 
where $J_\al$ is a projected current defined as 
\begin{align}
J_\al \equiv \frac{1}{1-2[R_{\rm Jor}]_g\circ P_2}A_\al \,. 
\end{align}
Here the parameter $\eta$ in \eqref{action} is set as $\eta=1$\,. 
The linear R-operator $R_{\rm Jor}$ associated with \eqref{Jor-r} is represented 
by the identification \eqref{linearR}. 

\medskip 

To find the AdS$_5$ part of \eqref{LGLB}, we use the following parameterization,  
\begin{align}
g=\exp\left(p_0x^0+p_1x^1+p_2x^2+p_3x^3\right)\exp\left(\frac{\ga_5}{2}\log z\right)
\quad \in SU(2,2)\,. 
\label{g-param}
\end{align}
For the $\alg{su}(2,2)$ generators $p_\mu$ ($\mu=0,1,2,3$) and $\ga_5$, 
see Appendix \ref{app:notation}.  
The projected deformed current $P_2(J_\al)$ is obtained by solving the equation,  
\begin{align}
(1-2P_2\circ R_g)P_2(J_\al)=P_2(A_\al)\,. 
\label{rel}
\end{align}
By plugging the projected current\footnote{%
For the convention of $\ga$-matrices, see Appendix \ref{app:notation}.}  
\begin{align}
P_2(A_\al) = \frac{\partial_\al x^0 \ga_0+\partial_\al x^1 \ga_1+\partial_\al x^2\ga_2
+\partial_\al x^3 \ga_3+\partial_\al z \ga_5}{2z}  
\end{align}
with \eqref{rel}, the deformed current is evaluated as
\begin{align}
P_2(J_\al)=j_\al^0\ga_0+j_\al^1\ga_1+j_\al^2\ga_2+j_\al^3\ga_3+j_\al^5\ga_5\,, 
\end{align}
with the coefficients 
\begin{align}
j_\al^1&= \frac{1}{2z}\partial_\al x^1
-\frac{(c_1+c_2)x^1+i (c_1-c_2)x^2}{2\sqrt{2}z^3}\partial_\al x^+ \,, \el \\ 
j_\al^2&= \frac{1}{2z}\partial_\al x^2  
-\frac{(c_1+c_2)x^2-i(c_1-c_2)x^1}{2\sqrt{2} z^3}\partial_\al x^+  \,,\el \\
j_\al^0+j_\al^3&=\frac{1}{\sqrt{2}z}\partial_\al x^+ \,, \el \\ 
j_\al^0-j_\al^3&=
-\frac{(c_1+c_2)x^1+i(c_1-c_2)x^2}{2z^3}\partial_\al x^1 
-\frac{(c_1+c_2)x^2-i (c_1-c_2)x^1}{2 z^3}\partial_\al x^2 \el \\
&\quad 
+\frac{4c_1c_2 ((x^1)^2+(x^2)^2)+(c_1+c_2)^2z^2 }{2\sqrt{2} z^5}\partial_\al x^+
+\frac{1}{\sqrt{2} z}\partial_\al x^-
-\frac{c_1+c_2}{2 z^2}\partial_\al z\,, \el \\
j_\al^5&=
\frac{1}{2z}\partial_\al z -\frac{c_1+c_2}{2\sqrt{2} z^2}\partial_\al x^+\,.   
\end{align}
Here we have introduced the light-cone coordinates
\begin{align}
x^\pm \equiv \frac{x^0\pm x^3}{\sqrt{2}}\,. 
\end{align}
Finally, the metric part and NS-NS two-form part of the Lagrangian \eqref{LGLB}
are given by 
\begin{align}
L_G&=-\ga^{\al\be}\Bigl[\frac{-2\partial_\al x^+\partial_\be x^- +\partial_\al x^1\partial_\be x^1
+\partial_\al x^2\partial_\be x^2+\partial_\al z\partial_\be z}{2 z^2} \el \\
&\qquad\qquad 
-\frac{4c_1c_2 ((x^1)^2 + (x^2)^2)+(c_1+c_2)^2z^2 }{8 z^6}\partial_\al x^+\partial_\be x^+\Bigr] \,, 
\label{LGLB-final}\\
L_B&=-\ep^{\al\be}\Bigl[(c_1+c_2)\frac{ (x^1\partial_\al x^1 + x^2 \partial_\al x^2 + z\partial_\al z)\partial_\be x^+}{2 z^4} 
+i(c_1-c_2 )\frac{(x^2\partial_\al x^1 - x^1\partial_\al x^2)\partial_\be x^+}{2 z^4}\Bigr]\,. \el 
\end{align}
Note that the resulting Lagrangian becomes complex in general. 
Therefore it is necessary to argue the reality condition.

\subsubsection*{Reality condition} 

Interestingly, both $L_G$ and $L_B$ become real if and only if $c_1$ and $c_2$ are related by the complex conjugation, 
\begin{align}
c_1 = c_2^{\ast} \qquad \Longleftrightarrow \qquad 
c_1=\alpha\,, \quad c_2=\alpha^* \qquad (\alpha \in \mathbb{C})\,. 
\label{realcon}
\end{align}
In particular, the result given in Subsec.\ 2.2 of \cite{SUGRA-KMY} can be reproduced by imposing that 
\begin{align}
c_1=c_2=\frac{1}{\sqrt{2}}\,. 
\end{align}

\subsection{The metric and NS-NS two-form}
\label{subsec:metric}

In the previous subsection, we have seen that a Jordanian $r$-matrix \eqref{Jor-r} yields 
the metric in the string frame and NS-NS two form, as presented in \eqref{LGLB-final}. 
Taking into account the reality condition \eqref{realcon}, 
the associated metric and NS-NS two form are derived as   
\begin{align}
ds^2&=
\frac{-2dx^+dx^- + (dx^1)^2 + (dx^2)^2+dz^2}{z^2}
-\frac{|\alpha|^2 \bigl( (x^1)^2 + (x^2)^2 \bigr)+ \bigl({\rm Re}(\alpha)\bigr)^2z^2 }{z^6}(dx^+)^2 \,, 
\nonumber  \\ 
B_2&=-{\rm Re}(\alpha) \frac{ (x^1 dx^1 + x^2 dx^2 + zdz)\wedge dx^+}{z^4} 
+{\rm Im}(\alpha) \frac{(x^2 dx^1 - x^1 dx^2)\wedge dx^+}{z^4} \,.  \label{ga-def}
\end{align}
It is worth seeing two special cases of $\alpha$ as listed below.

\subsubsection*{(i) Pure imaginary $\alpha$-deformation} 

When $\alpha$ is pure imaginary,
\begin{align}
\alpha \equiv i\alpha_{\rm I} \qquad \text{with} \qquad \alpha_{\rm I} \in \mathbb{R}\,, 
\end{align}
the metric and NS-NS two-form in \eqref{ga-def} are given by 
\begin{align}
ds^2&=
\frac{-2dx^+dx^- + (dx^1)^2 + (dx^2)^2+dz^2}{z^2}
- \alpha_{\rm I}^2 \frac{(x^1)^2 + (x^2)^2}{z^6}(dx^+)^2 \,, 
\nln
B_2&= \alpha_{\rm I}\, \frac{(x^2 dx^1 - x^1 dx^2)\wedge dx^+}{z^4} \,.  
\label{img-ga}
\end{align}
These agree with the ones of the solution found in \cite{HRR}.

\subsubsection*{(ii) Real $\alpha$-deformation} 

When $\alpha$ is a real number,  
\begin{align}
\alpha = \alpha_{\rm R} \qquad \text{with} \qquad \alpha_{\rm R} \in \mathbb{R}\,, 
\end{align}
the metric and NS-NS two-form in \eqref{ga-def} is rewritten as  
\begin{align}
ds^2&=\frac{-2dx^+dx^- + (dx^1)^2 + (dx^2)^2+dz^2}{z^2} 
- \alpha_{\rm R}^2 \frac{(x^1)^2 + (x^2)^2+z^2 }{z^6}(dx^+)^2\,, 
\nln
B_2&= -\alpha_{\rm R} \frac{ (x^1 dx^1 + x^2 dx^2+zdz)\wedge dx^+}{z^4} \,.  
\label{real-ga}
\end{align}
The above metric and NS-NS two-form are nothing but the ones obtained in \cite{SUGRA-KMY}.

\section{A chain of dualities for AdS$_5\times$S$^5$} 
\label{sec:chain}

In this section, we show that the metric and NS-NS two-form in (\ref{ga-def}) 
are reproduced by performing TsT-transformations and an S-duality for the AdS$_5\times$S$^5$ background.  

\medskip 

In order to perform T-dualities, let us introduce 
the polar coordinates on the $x^1$-$x^2$ plane,  
\begin{align}
x^1= \rho \cos\varphi\,, \qquad x^2= \rho \sin\varphi ~~~\qquad (0 \leq \rho < \infty\,,
~~0\leq \varphi <2\pi)\,. 
\label{coord-trans}
\end{align}
With these coordinates, the original AdS$_5\times$S$^5$ background is given by  
\begin{eqnarray}
ds^2 &=& \frac{-2dx^+dx^- +d\rho^2+\rho^2 d\varphi^2 +dz^2}{z^2} + (d\chi + \omega)^2 +ds^2_{\mathbb{C}{\rm P}^2} \,, 
\label{AdS} \\ 
F_5 &=& dC_4 = 4 \biggl[
-\frac{1}{z^5}\,dx^+\wedge dx^-\wedge d\rho \wedge \rho d\varphi \wedge dz  \nonumber \\
&& + (d\chi+\omega)\wedge d\mu \wedge \sin\mu\,\Sigma_1\wedge \sin\mu\,\Sigma_2\wedge \cos\mu\sin\mu\,\Sigma_3
\biggr]\,,     
\nonumber \\ 
B_2 &=& C_2 = C = 0\,, \qquad \Phi = \Phi_0~~(\mbox{const.})\,. \nonumber 
\end{eqnarray}
Note that the metric of S$^5$ is expressed as a $U(1)$-fibration over $\mathbb{C}$P$^2$\,.  
Here $\chi$ is a local coordinate on the fiber and $\omega$ is a one-form potential for the K\"ahler form 
on $\mathbb{C}$P$^2$\,. In the usual way, the metric of $\mathbb{C}{\rm P}^2$ and $\omega$ are given by 
\begin{eqnarray}
ds^2_{\mathbb{C}{\rm P}^2} = d\mu^2 +\sin^2\mu\left(\Sigma_1^2+\Sigma_2^2 + \cos^2\mu\,\Sigma_3^2\right)\,, 
\qquad \omega \equiv \sin^2\mu\,\Sigma_3\,,
\end{eqnarray}
where $\Sigma_a~(a=1,2,3)$ are defined as 
\begin{eqnarray}
&& \Sigma_1 \equiv \frac{1}{2}\left(\cos\psi\,d\theta + \sin\psi\sin\theta\,d\phi\right)\,, \quad 
\Sigma_2 \equiv \frac{1}{2}\left(\sin\psi\,d\theta - \cos\psi\sin\theta\,d\phi\right)\,, \nonumber \\ 
&& \Sigma_3 \equiv \frac{1}{2}\left(d\psi + \cos\theta\,d\phi\right)\,. \nonumber 
\end{eqnarray}
The geometry of S$^5$ is described with the five coordinates: $(\chi, ~\mu,~ \psi,~\theta,~\phi)$\,.  

\medskip 

For later computations, it is nice to write down explicitly the R-R four-form $C_4$\,, 
\begin{eqnarray}
C_4 &=& \frac{1}{z^4}\,dx^+\wedge dx^-\wedge d\rho \wedge \rho d\varphi 
-\sin^4\mu\,d\chi\wedge \Sigma_1\wedge \Sigma_2 \wedge \Sigma_3\,. 
\end{eqnarray}
Note that the following relations are satisfied, 
\[
\Sigma_1\wedge \Sigma_2 \wedge \Sigma_3 = -\frac{1}{8}\sin\theta\,d\theta \wedge d\phi \wedge d\psi\,, 
\qquad d(\Sigma_1\wedge \Sigma_2 \wedge \Sigma_3) =0\,. 
\]
It is also helpful to use the relations, 
\[
dx^1\wedge dx^2 = d\rho\wedge \rho d\varphi\,, \quad 
x^1dx^1 + x^2 dx^2 = \rho d\rho\,, \quad x^2dx^1-x^1 dx^2 = -\rho^2 d\varphi\,.
\]

\medskip 

In the following, we will apply TsT transformations and an S-duality 
for the AdS$_5\times$S$^5$ background in (\ref{AdS})\,.

\subsection{The first TsT transformation}
\label{subsec:1stTsT}

The first step is to perform a TsT transformation to the AdS$_5\times$S$^5$ background in (\ref{AdS})\,. 

\medskip 

Let us first take a T-duality along the $\varphi$-direction in (\ref{AdS})\,. 
According to the rules of T-duality (listed in Appendix \ref{app:Tdual}), 
the background is rewritten as  
\begin{align}
d\tilde{s}^2&=\frac{-2dx^+dx^- +d\rho^2 +dz^2}{z^2}+\frac{z^2}{\rho^2}d\tilde{\varphi}^2 +ds^2_{\mathrm{S}^5} \,,  \nonumber \\
\tilde{B}_2&=0\,,\qquad \tilde \Phi=\Phi_0-\frac{1}{2}\ln \left(\frac{\rho^2}{z^2}\right)\,, \nonumber \\ 
\tilde{C}_3 &= -\frac{\rho}{z^4}\,dx^+\wedge dx^-\wedge d\rho\,, \qquad 
\tilde{C}_5 = -\sin^4 \mu\, d\chi\wedge \Sigma_1\wedge \Sigma_2 \wedge \Sigma_3 \wedge d\tilde{\varphi}\,.
\end{align}
Then the $x^-$-coordinate is shifted as  
\begin{align}
x^-\to x^- + a_1\, \tilde{\varphi} 
\end{align}
with a real parameter $a_1$\,. The metric and $\tilde{C}_3$ are deformed as follows:  
\begin{align}
d\tilde{s}^2&=\frac{-2dx^+dx^- +d\rho^2 +dz^2}{z^2} - 2\frac{a_1}{z^2}dx^+d\tilde{\varphi} 
+\frac{z^2}{\rho^2}d\tilde{\varphi}^2 +ds^2_{\mathrm{S}^5}\,, \nonumber  \\
\tilde B_2&=0\,, \qquad \tilde \Phi=\Phi_0-\frac{1}{2}\ln \left(\frac{\rho^2}{z^2}\right)\,, \nonumber \\ 
\tilde{C}_3 &=  -\frac{\rho}{z^4}\,dx^+ \wedge dx^-\wedge d\rho - a_1 \frac{\rho}{z^4}\,dx^+\wedge d\tilde{\varphi}\wedge d\rho \,, \nonumber \\  
\tilde{C}_5 &= -\sin^4 \mu\, d\chi\wedge \Sigma_1\wedge \Sigma_2 \wedge \Sigma_3 \wedge d\tilde{\varphi}\,. 
\end{align}
Finally, by taking a T-duality along the $\tilde{\varphi}$-direction, the resulting background is given by 
\begin{align}
ds^2&=\frac{-2dx^+dx^- +d\rho^2+\rho^2 d\varphi^2 +dz^2}{z^2} 
- a_1^2 \frac{\rho^2}{z^6}(dx^+)^2 +ds^2_{\mathrm{S}^5}\,, \nonumber \\
B_2&= -a_1 \frac{\rho^2}{z^4}dx^+\wedge d\varphi\,, \qquad \Phi=\Phi_0\,, \qquad 
C_2 = a_1 \frac{\rho}{z^4}\,dx^+\wedge d\rho\,, \nonumber \\  
C_4 &= \frac{\rho}{z^4}\,dx^+\wedge dx^- \wedge d\rho\wedge d\varphi - \sin^4 \mu\, d\chi\wedge \Sigma_1\wedge \Sigma_2 \wedge \Sigma_3\,. 
\label{HRR}    
\end{align}
Note that the dilaton is constant. 
Through the coordinate transformations \eqref{coord-trans}, 
the above metric and NS-NS two-form agree with the ones in \eqref{img-ga} 
when $a_1=-{\rm Im}(\al)=-\alpha_{\rm I}$\,. 
The NS-NS sector of the background (\ref{HRR}) has already been obtained in \cite{HRR}.

\subsection{The second TsT transformation}

The next step is to perform another TsT transformation for the background (\ref{HRR})\,. 
This process is essentially the same as the one in \cite{MMT}. 

\medskip 

Let us first take a T-duality along the $\chi$-direction of S$^5$\,. The resulting metric is given by 
\begin{align}
ds^2&=\frac{-2dx^+dx^- +d\rho^2+\rho^2 d\varphi^2 +dz^2}{z^2} 
-a_1^2 \frac{\rho^2}{z^6}(dx^+)^2 + d\tilde{\chi}^2 +ds^2_{\mathbb{C}{\rm P}^2}\,, \nonumber \\
\tilde{B}_2& = -a_1 \frac{\rho^2}{z^4}dx^+\wedge d\varphi + \frac{1}{2}\sin^2\mu\,d\psi\wedge d\tilde{\chi} 
+ \frac{1}{2}\sin^2\mu\cos\theta\,d\phi\wedge d\tilde{\chi}\,, \qquad \Phi=\Phi_0\,, \nonumber \\ 
\tilde{C}_3 &=  a_1 \frac{\rho}{z^4}\,dx^+\wedge d\rho \wedge d\tilde{\chi} 
- \sin^4 \mu\, \Sigma_1\wedge \Sigma_2 \wedge \Sigma_3\,, \nonumber \\  
\tilde{C}_5 &= \frac{\rho}{z^4}\,dx^+\wedge dx^- \wedge d\rho\wedge d\varphi\wedge d\tilde{\chi}\,. 
\end{align}
Then, by shifting $x^-$ as 
\begin{eqnarray}
x^- \to x^- +a_2\, \tilde{\chi}\,, 
\end{eqnarray} 
only the metric is deformed as 
\begin{eqnarray}
ds^2 =\frac{-2dx^+dx^- +d\rho^2+\rho^2 d\varphi^2 +dz^2}{z^2} 
-a_1^2 \frac{\rho^2}{z^6}(dx^+)^2 -2\frac{a_2}{z^2}dx^+d\tilde{\chi} 
+ d\tilde{\chi}^2 +ds^2_{\mathbb{C}{\rm P}^2}\,. \nonumber 
\end{eqnarray}
Finally, by taking a T-duality along the $\tilde{\chi}$-direction, 
the resulting background is given by 
\begin{align}
ds^2&=\frac{-2dx^+dx^- +d\rho^2+\rho^2 d\varphi^2 +dz^2}{z^2} 
-\left(a_1^2 \frac{\rho^2}{z^6} + a_2^2 \frac{1}{z^4}\right)(dx^+)^2
+ (d\chi +\omega)^2 +ds^2_{\mathbb{C}{\rm P}^2}\,, \nonumber \\
B_2& = -a_1 \frac{\rho^2}{z^4}dx^+\wedge d\varphi -\frac{a_2}{z^2}dx^+\wedge (d\chi+\omega)\,, 
\qquad \Phi=\Phi_0\,, \qquad C_2 =  a_1 \frac{\rho}{z^4}\,dx^+\wedge d\rho\,,  \nonumber \\  
C_4 &= \frac{\rho}{z^4}\,dx^+\wedge dx^- \wedge d\rho\wedge d\varphi 
- \sin^4 \mu\,d\chi\wedge \Sigma_1\wedge \Sigma_2 \wedge \Sigma_3\,. 
\label{sol2}
\end{align}
As a result, a deformation term of the Schr\"odinger spacetime has been added. 

\medskip 

When $a_1=0$\,, the Schr\"odinger solution is realized \cite{MMT}. 
In comparison to the $r$-matrix (\ref{Jor-r}) that is composed of $\mathfrak{su}(2,2)$ only,  
the classical $r$-matrix associated with this solution itself is 
composed of both $\mathfrak{su}(2,2)$ and $\mathfrak{su}(4)$ generators 
because the TsT transformation includes a direction of S$^5$\,.  
We will explain in very detail the $r$-matrix corresponding to the Schr\"odinger spacetime in another place \cite{future}. 
Notably, this result indicates that the Schr\"odinger spacetime is integrable\footnote{
The integrability of Schr\"odinger spacetimes may be related to the coset structure argued in \cite{SYY}. 
A classification of super Schr\"odinger algebras \cite{SY} would also play an important role in the future.}.
A further remarkable point is that brane-wave type deformations \cite{HY} also lead to integrable backgrounds.

\subsection{S-duality}

Then let us perform an S-duality for the background (\ref{sol2})\,. 

\medskip 

The transformation rule of S-duality is given by 
\begin{eqnarray}
\Phi' = -\Phi\,, \qquad B_2' = C_2\,, \qquad C_2' = -B_2\,.
\label{S-dual}
\end{eqnarray}
Hence, after performing the S-duality, the resulting background is given by 
\begin{align}
ds^2&=\frac{-2dx^+dx^- +d\rho^2+\rho^2 d\varphi^2 +dz^2}{z^2} 
-\left(a_1^2 \frac{\rho^2}{z^6} + a_2^2 \frac{1}{z^4}\right)(dx^+)^2
+ (d\chi +\omega)^2 +ds^2_{\mathbb{C}{\rm P}^2}\,, \nonumber \\
B_2 &=  a_1 \frac{\rho}{z^4}\,dx^+\wedge d\rho\,, \qquad \Phi=-\Phi_0\,, \qquad 
C_2 = a_1 \frac{\rho^2}{z^4}dx^+\wedge d\varphi 
+\frac{a_2}{z^2}dx^+\wedge (d\chi+\omega)\,, 
\nonumber \\  
C_4 &= \frac{\rho}{z^4}\,dx^+\wedge dx^- \wedge d\rho\wedge d\varphi 
- \sin^4 \mu\, d\chi\wedge\Sigma_1\wedge \Sigma_2 \wedge \Sigma_3\,. 
\label{us}
\end{align}
When $a_1=\pm a_2={\rm Re}(\al)=\al_{\rm R}$\,, 
the above background agrees with the one in \eqref{real-ga}\,. 
This solution is nothing but the one obtained in \cite{SUGRA-KMY}, 
where the deformation parameter was denoted by $\al_{\rm R}\equiv \eta$\,. 
Notably, this result ensures that the solution in \cite{SUGRA-KMY} is a consistent string background. 
In particular, the world-sheet beta-function vanishes.

\subsection{The third TsT transformation}

Furthermore, let us perform a TsT-transformation, which is the same as in Sec.~\ref{subsec:1stTsT}\,, 
for the background in (\ref{us}). 

\medskip 

The background in (\ref{us}) seems complicated, but it has a $U(1)$ symmetry along the $\varphi$-direction. 
We first take a T-duality along the $\varphi$-direction. As a result, the background is transformed as  
\begin{align}
ds^2&=\frac{-2dx^+dx^- +d\rho^2 +dz^2}{z^2} + \frac{z^2}{\rho^2} d\tilde{\varphi}^2
-\left(a_1^2 \frac{\rho^2}{z^6} + a_2^2 \frac{1}{z^4}\right)(dx^+)^2
+ (d\chi +\omega)^2 +ds^2_{\mathbb{C}{\rm P}^2}\,, \nonumber \\
B_2 &=  a_1 \frac{\rho}{z^4}\,dx^+\wedge d\rho\,, \qquad 
\Phi=-\Phi_0 - \frac{1}{2}\ln\left(\frac{\rho^2}{z^2}\right)\,, 
\qquad C_1 =  -a_1 \frac{\rho^2}{z^4}dx^+\,, \nonumber \\ 
C_3 &= \frac{a_2}{z^2}dx^+\wedge (d\chi+\omega)\wedge d\tilde{\varphi} 
- \frac{\rho}{z^4}\,dx^+\wedge dx^- \wedge d\rho\,, 
\nonumber \\  
C_5 &= - \sin^4 \mu\, d\chi\wedge\Sigma_1\wedge \Sigma_2 \wedge \Sigma_3\wedge d\tilde{\varphi}\,. 
\end{align}
Then, by shifting $x^-$ as 
\begin{eqnarray}
x^- \to x^- +a_3\,\tilde{\varphi}\,, 
\end{eqnarray}
it is rewritten into the following form: 
\begin{align}
ds^2&=\frac{-2dx^+dx^- +d\rho^2 +dz^2}{z^2} -2\frac{a_3}{z^2}dx^+d\tilde{\varphi}
+ \frac{z^2}{\rho^2} d\tilde{\varphi}^2
-\left(a_1^2 \frac{\rho^2}{z^6} + a_2^2 \frac{1}{z^4}\right)(dx^+)^2 \nonumber \\ 
& \qquad + (d\chi +\omega)^2 +ds^2_{\mathbb{C}{\rm P}^2}\,, \nonumber \\
B_2 &=  a_1 \frac{\rho}{z^4}\,dx^+\wedge d\rho\,, \qquad \Phi=-\Phi_0 - \frac{1}{2}\ln\left(\frac{\rho^2}{z^2}\right)\,, 
\qquad C_1 =  -a_1 \frac{\rho^2}{z^4}dx^+\,, \nonumber \\ 
C_3 &= \frac{a_2}{z^2}dx^+\wedge (d\chi+\omega)\wedge d\tilde{\varphi} 
-\frac{\rho}{z^4}\,dx^+\wedge dx^- \wedge d\rho
+a_3\frac{\rho}{z^4}\,dx^+ \wedge d\rho \wedge d\tilde{\varphi}\,, \nonumber \\  
C_5 &= - \sin^4 \mu\, d\chi\wedge\Sigma_1\wedge \Sigma_2 \wedge \Sigma_3\wedge d\tilde{\varphi}\,. 
\end{align}
Finally, by taking a T-duality along the $\tilde{\varphi}$-direction,  
the resulting background is given by 
\begin{align}
ds^2&=\frac{-2dx^+dx^- +d\rho^2 + \rho^2 d\varphi^2 +dz^2}{z^2} 
-\left( (a_1^2 +a_3^2) \frac{\rho^2}{z^6} + a_2^2 \frac{1}{z^4}\right)(dx^+)^2 \nonumber \\ 
& \qquad + (d\chi +\omega)^2 +ds^2_{\mathbb{C}{\rm P}^2}\,, \nonumber \\
B_2 &=  a_1 \frac{\rho}{z^4}\,dx^+\wedge d\rho - a_3\frac{\rho^2}{z^4}dx^+\wedge d\varphi\,, 
\qquad \Phi=-\Phi_0\,, \nonumber \\ 
C_2 &= a_1 \frac{\rho^2}{z^4}dx^+ \wedge d\varphi  
+\frac{a_2}{z^2}dx^+\wedge (d\chi+\omega) 
+ a_3\frac{\rho}{z^4}\,dx^+ \wedge d\rho \,, \nonumber \\ 
C_4 &=  \frac{\rho}{z^4}\,dx^+\wedge dx^- \wedge d\rho \wedge d\varphi 
- \sin^4 \mu\, d\chi\wedge\Sigma_1\wedge \Sigma_2 \wedge \Sigma_3\,. 
\label{final}
\end{align}
When $a_1= \pm a_2 = \mathrm{Re}(\alpha)$ and 
$a_3 = -\mathrm{Im}(\alpha)$\,, the solution (\ref{final}) reproduces 
the metric and NS-NS two-form in (\ref{ga-def}). Note that in this identification 
the sign concerning $a_2$ has not been determined here. In order to fix this ambiguity, 
we have to perform a supercoset construction. 

\medskip 

So far, the R-R sector has not been determined yet from the Yang-Baxter sigma model approach. 
But the solution (\ref{final}) gives a prediction for the R-R sector. 
If the gravity/CYBE correspondence is true, then the R-R sector of (\ref{final}) should be reproduced 
by performing a supercoset construction. We will not try to do that here and leave it as a future problem. 
However, it should be derived as expected. The kappa-invariance of the deformed string action would also imply the agreement.

\section{Duality-chains and classical $r$-matrices }
\label{sec:summary} 

It is worth summarizing the relation between duality-chains and classical $r$-matrices. 
This is nothing but the gravity/CYBE correspondence concerned with the present scope. 

\medskip 

For the completeness, we first consider the fourth TsT-transformation
in subsection \ref{subsec:TsT4} and the second S-duality in 
subsection \ref{subsec:2nd-S}\,.  
In subsection \ref{subsec:r-3para}, we propose a classical $r$-matrix including 
three parameters and summarize a concrete realization of 
the gravity/CYBE correspondence.

\subsection{The fourth TsT transformation} 
\label{subsec:TsT4}

Let us consider here a TsT-transformation for the background in \eqref{final}\,. 

\medskip 

We first perform a T-duality for the background in \eqref{final} along the $\chi$-direction. 
The resulting background is 
\begin{align}
ds^2&=ds^2_{\rm AdS_5} 
-\left( (a_1^2 +a_3^2) \frac{\rho^2}{z^6} + a_2^2 \frac{1}{z^4}\right)(dx^+)^2 
+ d\tilde\chi^2 +ds^2_{\mathbb{C}{\rm P}^2}\,, \nonumber \\
B_2 &=  dx^+\wedge \left(a_1 \frac{\rho d\rho}{z^4}  
- a_3\frac{\rho^2d\varphi}{z^4} \right)
+\omega \wedge d\tilde\chi \,, 
\qquad \tilde\Phi=-\Phi_0\,, 
\qquad C_1=-\frac{a_2}{z^2}dx^+\,, 
\nln
C_3 &= dx^+\wedge\left(a_1 \frac{\rho^2 d\varphi}{z^4}  
+ a_3\frac{\rho d\rho}{z^4}\right)\wedge \tilde d\chi
-\sin^4\mu\, \Sigma_1\wedge\Sigma_2\wedge\Sigma_3\,,   \nln
C_5&=\frac{\rho}{z^4}dx^+\wedge dx^-\wedge d\rho\wedge d\phi\wedge d\tilde\chi\,, 
\label{TsT4-1}
\end{align}
where the undeformed metric of AdS$_5$ is denoted as 
\begin{align}
ds^2_{\rm AdS_5}=\frac{-2dx^+dx^- +d\rho^2 + \rho^2 d\varphi^2 +dz^2}{z^2} \,. 
\end{align}
Then, by shifting the $x^-$-coordinate as 
\begin{eqnarray}
x^- \to x^- + a_4\, \tilde\chi\,, 
\end{eqnarray}
only the metric in \eqref{TsT4-1} is modified as 
\begin{align}
ds^2&= ds^2_{\rm AdS_5} -\frac{2a_4}{z^2}dx^+d\tilde\chi
-\left( (a_1^2 +a_3^2) \frac{\rho^2}{z^6} + a_2^2 \frac{1}{z^4}\right)(dx^+)^2
+ d\tilde\chi^2 +ds^2_{\mathbb{C}{\rm P}^2} \,. 
\end{align}
Finally, by T-dualizing back along the $\tilde\chi$-direction,  
we arrive at the following background: 
\begin{align}
ds^2&=ds^2_{\rm AdS_5}
-\left( (a_1^2 +a_3^2) \frac{\rho^2}{z^6} + (a_2^2 +a_4^2)\frac{1}{z^4}\right)(dx^+)^2 
+ (d\chi+\omega)^2 +ds^2_{\mathbb{C}{\rm P}^2}\,, \nonumber \\
B_2 &=  dx^+\wedge \left(
-a_3\frac{\rho^2d\varphi}{z^4} 
-a_4\frac{d\chi+\omega}{z^2} 
+a_1 \frac{\rho d\rho}{z^4}  \right)\,, 
\qquad \Phi=-\Phi_0\,, \nln
C_2&=dx^+\wedge \left(
a_1\frac{\rho^2d\varphi}{z^4} 
+a_2\frac{d\chi+\omega}{z^2} 
+a_3 \frac{\rho d\rho}{z^4}  \right)\,,  \nln
C_4&= \frac{\rho}{z^4}\,dx^+\wedge dx^- \wedge d\rho \wedge d\varphi 
- \sin^4 \mu\, d\chi\wedge\Sigma_1\wedge \Sigma_2 \wedge \Sigma_3\,. 
\label{TsT4}
\end{align}

\subsection{The second S-duality } 
\label{subsec:2nd-S}

One may notice that, in the background \eqref{TsT4}\,, 
the NS-NS two-form $B_2$ and the Ramond-Ramond two-form $C_2$ are turned on in  
a symmetric way. This is because the background \eqref{TsT4} is obtained by two same 
TsT-transformations before and after the S-duality \eqref{S-dual}\,. 
Then, let us consider the second S-duality of \eqref{TsT4}\,. 
The resulting background is given by 
\begin{align}
ds^2&=ds^2_{\rm AdS_5}
-\left( (a_1^2 +a_3^2) \frac{\rho^2}{z^6} + (a_2^2 +a_4^2)\frac{1}{z^4}\right)(dx^+)^2 
+ (d\chi+\omega)^2 +ds^2_{\mathbb{C}{\rm P}^2}\,, \nonumber \\
B_2&=dx^+\wedge \left(
a_1\frac{\rho^2d\varphi}{z^4} 
+a_2\frac{d\chi+\omega}{z^2} 
+a_3 \frac{\rho d\rho}{z^4}  \right)\,,  
\qquad \Phi=\Phi_0\,, \nln
C_2 &= dx^+\wedge \left(
a_3\frac{\rho^2d\varphi}{z^4} 
+a_4\frac{d\chi+\omega}{z^2} 
-a_1 \frac{\rho d\rho}{z^4}  \right)\,, \nln
C_4&= \frac{\rho}{z^4}\,dx^+\wedge dx^- \wedge d\rho \wedge d\varphi 
- \sin^4 \mu\, d\chi\wedge\Sigma_1\wedge \Sigma_2 \wedge \Sigma_3\,. 
\label{TsT4-S}
\end{align}

\subsection{A three-parameter deformation of AdS$_5\times$S$^5$  } 
\label{subsec:r-3para}

Now it is turn to consider a classical $r$-matrix corresponding to the background (\ref{TsT4-S}). 
It is argued to be the following form, 
\begin{align}
r_{12}(b_1,b_2,b_3)=\frac{1}{\sqrt{2}i}E_{24}\wedge \left(
b_1(E_{22}+E_{44})+\frac{b_2}{2}(h_4+h_5+h_6)+ib_3(E_{22}-E_{44})\right)\,.
\label{r-3para}
\end{align}
Here $h_4,h_5$ and $h_6$ are the three Cartan generators\footnote{
The convention of the $\alg{su}(4)$ algebra would be presented in \cite{future}\,. The case with $b_1=b_3=0$ 
will be elaborated in \cite{future}.} 
of $\alg{su}(4)$ 
rather than $\alg{su}(2,2)$\,, and   
$b_1,b_2,b_3$ are real deformation parameters. 
A direct computation shows that this is a solution of the CYBE. 

\medskip 

Plugging the classical $r$-matrix \eqref{r-3para} with the classical action \eqref{action}\,, 
the resulting metric and NS-NS two-form turn out to be 
\begin{align}
ds^2&=ds^2_{\rm AdS_5}+ds^2_{\rm S^5}
-\frac{(b_1^2+b_3^2)\rho^2+(b_2^2+b_3^2)z^2}{z^6}(dx^+)^2\,, \\
B_2&= dx^+\wedge \left(
-b_1\frac{\rho^2d\varphi}{z^4}-b_2\frac{d\chi+\omega}{z^2}+b_3\frac{\rho d\rho}{z^4}
\right)\,,  
\end{align}
where $ds^2_{\rm AdS_5}$ and $ds^2_{\rm S^5}$ are the undeformed metrics 
of AdS$_5$ and S$^5$\,, respectively. 

\medskip 

Indeed, the above metric and NS-NS two-form agree with the ones obtained by the fourth 
TsT-transformation in \eqref{TsT4} and the S-duality in \eqref{TsT4-S} 
with the following parameter identifications, respectively,   
\begin{align}
&b_1= a_3\,, &&b_2= a_4\,,  &&b_3= a_1=\pm a_2 &&\text{for} \qquad \eqref{TsT4}\,,  \\ 
&b_1= -a_1\,,&&b_2= -a_2\,, &&b_3= a_3=\pm a_4 &&\text{for} \qquad \eqref{TsT4-S}\,.   
\end{align}

\medskip 

It is worth comparing the above results with the deformed backgrounds from TsT-transformations 
and an S-duality in Sec.\ \ref{sec:chain}\,. 
The comprehensive relations are summarized in Tab.\ \ref{list:tab}. 
Here the symbol (TsT)$^{a_1}_\varphi$\,, for instance,  
stands for a TsT-transformation consisting of a T-duality for the  
$\varphi$-direction and the shift $x^-\to x^-+a_1\tilde\varphi$\,. 
The capital S denotes an S-duality. For the details, see Sec.\ \ref{sec:chain}\,. 

\begin{table}[t]
\begin{center}
\begin{tabular}{|l|r|l|} 
\hline 
Classical $r$-matrices & Duality chains& \quad Backgrounds 
\\ \hline\hline  
$r_{12}(a_1,\,0\,,\,0\,)$  & (TsT)$^{a_1}_\varphi$ & \eqref{HRR}\,, see also \cite{HRR}
\\ 
$r_{12}(a_1,a_2,\,0\,)$  & 
(TsT)$^{a_2}_\chi$$\circ$(TsT)$^{a_1}_\varphi$ & \eqref{sol2}
\\ 
$r_{12}(\,0\,,\,0\,,a_1)$ & S$\circ$(TsT)$^{\pm a_1}_\chi$$\circ$(TsT)$^{a_1}_\varphi$ 
& \eqref{us}\,, see also \cite{SUGRA-KMY}
\\ 
$r_{12}(a_3,\,0\,,a_1)$ & (TsT)$^{a_3}_\varphi$$\circ$
S$\circ$(TsT)$^{\pm a_1}_\chi$$\circ$(TsT)$^{a_1}_\varphi$  & \eqref{final}
\\ 
$r_{12}(a_3,a_4,a_1)$ & (TsT)$^{a_4}_\chi$$\circ$(TsT)$^{a_3}_\varphi$$\circ$
S$\circ$(TsT)$^{\pm a_1}_\chi$$\circ$(TsT)$^{a_1}_\varphi$  & \eqref{TsT4}
\\ 
$r_{12}(-a_1,-a_2,a_3)$ & S$\circ$(TsT)$^{\pm a_3}_\chi$
$\circ$(TsT)$^{a_3}_\varphi$$\circ$
S$\circ$(TsT)$^{a_2}_\chi$$\circ$(TsT)$^{a_1}_\varphi$  & \eqref{TsT4-S}
\\\hline 
\end{tabular}
\end{center}
\caption{\footnotesize 
The classical $r$-matrix in \eqref{r-3para} and  
the associated duality chains. 
\label{list:tab}} 
\end{table}

\medskip

As for Tab.\ \ref{list:tab}, note that
the following duality chains including S-dualities 
are realized only if some parameter constraints are satisfied;   
\begin{align}
{\rm S}\circ({\rm TsT})^{a_2}_\chi \circ({\rm TsT})^{a_1}_\varphi 
\qquad\text{if}\qquad a_2=\pm a_1\,, 
\nln 
({\rm TsT})^{a_3}_\varphi \circ{\rm S}\circ({\rm TsT})^{a_2}_\chi \circ
({\rm TsT})^{a_1}_\varphi 
\qquad\text{if}\qquad a_2=\pm a_1\,,
\nln 
({\rm TsT})^{a_4}_\chi \circ({\rm TsT})^{a_3}_\varphi \circ{\rm S}
\circ({\rm TsT})^{a_2}_\chi \circ({\rm TsT})^{a_1}_\varphi 
\qquad\text{if}\qquad a_2=\pm a_1\,,
\nln 
{\rm S}\circ({\rm TsT})^{a_4}_\chi
\circ({\rm TsT})^{a_3}_\varphi\circ
{\rm S}\circ({\rm TsT})^{a_2}_\chi\circ({\rm TsT})^{a_1}_\varphi
\qquad\text{if}\qquad a_4=\pm a_3\,. 
\nonumber
\end{align}
In order to see the correspondence between the $r$-matrix in (\ref{r-3para}) 
and deformed backgrounds, we need to impose constraints for the values of $a_1, a_2, a_3, a_4$\,. 
It seems likely that the constraints would come from the S-duality, but at the present moment we have no idea for the origin.  
To reveal it, it would be important to develop Yang-Baxter deformations of 
the D-string action. Then the origin may be understood as a consistency condition 
of the S-duality transformation.

\section{Conclusion and discussion}
\label{sec:concl}

We have further studied integrable deformations of the
AdS$_5\times$S$^5$ superstring based on the Yang-Baxter sigma model approach
with classical $r$-matrices satisfying the CYBE. 
By focusing upon a classical $r$-matrix with two parameters and its three-parameter generalization, 
we have presented a family of the deformed metric and NS-NS two-form. 
The corresponding solutions of type IIB supergravity have been successfully obtained 
by performing TsT-transformations and S-dualities. 
In particular, it includes the solution found by Hubeny-Rangamani-Ross \cite{HRR} 
and the one found by us \cite{SUGRA-KMY} as special cases. 

\medskip 

So far, we have seen that a certain multi-parameter family of classical $r$-matrices correspond to 
a chain of TsT-transformations and S-dualities for the undeformed AdS$_5\times$S$^5$ 
background. It seems likely that Yang-Baxter sigma models based on CYBE 
could reproduce duality chains of AdS$_5\times$S$^5$\,. 
In fact, three-parameter $\gamma$-deformations of S$^5$ \cite{LM,Frolov} 
and gravity duals for noncommutative gauge theories \cite{HI,MR,DHH,MS} have been reproduced from 
the associated classical $r$-matrices \cite{LM-MY,MR-MY}. Integrable deformations 
based on (at least) a certain class of $r$-matrices can be undone with non-local gauge 
transformations and twisted boundary condition, by following the philosophy of \cite{AAF}. 

\medskip 

So far, we have discussed duality chains of AdS$_5\times$S$^5$\,. 
This is a simple story and one may think of that Yang-Baxter sigma models 
would work only for duality chains. However, this is not the case. 
For example, more complicated backgrounds have been presented in \cite{SUGRA-KMY} 
from the Yang-Baxter sigma model approach and those do not seem to be realized 
as TsT transformations of AdS$_5\times$S$^5$ because of the singular structure. 
Therefore, it seems likely that the Yang-Baxter sigma model approach would include 
a wider class of gravity solutions which cannot be realized by performing 
string dualities for the undeformed AdS$_5\times$S$^5$\,. 

\medskip 

It would be of great importance to clarify the applicability of Yang-Baxter sigma
models.  

\subsection*{Acknowledgments}

We are very grateful to Io Kawaguchi for collaboration at the very early stage of this work. 
We appreciate P.\ Marcos Crichigno, Minkyo Kim, Marc Magro and Benoit Vicedo 
for valuable comments and discussions. 
This work is supported in part by JSPS Japan-Hungary Research Cooperative Program.

\appendix

\section*{Appendix}

\section{Notation and convention }
\label{app:notation} 

An explicit basis of $\mathfrak{su}(2,2)$ is represented by  
the following $\gamma$-matrices, 
\begin{eqnarray}
&&\gamma_1=
{\small \begin{bmatrix}
~0~&~0~&~0~&-1\\
~0~&~0~&~1~&~0\\
~0~&~1~&~0~&~0\\
-1~&~0~&~0~&~0\\
\end{bmatrix}}\,, \quad
{\small \gamma_2=
\begin{bmatrix}
~0~&~0~&~0~&~i~\\
~0~&~0~&~i~&~0~\\
~0~&-i~&~0~&~0~\\
-i~&~0~&~0~&~0~\\
\end{bmatrix}}\,, \quad 
\gamma_3=
{\small \begin{bmatrix}
~0~&~0~&~1~&~0~\\
~0~&~0~&~0~&~1~\\
~1~&~0~&~0~&~0~\\
~0~&~1~&~0~&~0~\\
\end{bmatrix}}\,, \nonumber \\
&&\gamma_0=
{\small \begin{bmatrix}
~0~&~0~&1~&~0~\\
~0~&~0~&~0~&-1~\\
-1~&~0~&~0~&~0~\\
~0~&~1~&~0~&~0~\\
\end{bmatrix}}\,, \quad
\ga_5=i\ga_1\ga_2\ga_3\ga_0=   
{\small \begin{bmatrix}
~1~&~0~&~0~&~0~\\
~0~&~1~&~0~&~0~\\
~0~&~0~&-1~&~0~\\
~0~&~0~&~0~&-1~\\
\end{bmatrix}}\,. 
\end{eqnarray}
With the Lorentzian metric $\eta_{\mu\nu}={\diag}(-1,1,1,1)$\,, 
the Clifford algebra is satisfied as 
\begin{eqnarray}
\left\{\ga_\mu,\ga_\nu\right\}=2\eta_{\mu\nu}\,,\qquad 
\left\{\ga_\mu,\ga_5\right\}=0 \,, \qquad 
(\ga_5)^2=1\,. 
\end{eqnarray} 
The Lie algebra $\alg{so}(1,4)$ is formed by the generators 
\begin{align}
m_{\mu\nu}=\frac{1}{4}\left[\ga_\mu,\ga_\nu\right]\,, \qquad 
m_{\mu5}=\frac{1}{4}\left[\ga_\mu,\ga_5\right] \qquad 
(\mu,\nu=0,1,2,3\,)\,, 
\end{align}
and then enlarged algebra $\alg{so}(2,4)=\alg{su}(2,2)$ is spanned by the following set: 
\begin{align}
m_{\mu\nu}\,, \quad m_{\mu5}\,, \quad \frac{1}{2}\ga_\mu\,, \quad \frac{1}{2}\ga_5\,. 
\end{align}
The reality condition for these generators are given by 
\begin{align}
M^\dag \ga_0 + \ga_0 M=0 \qquad \text{for}\qquad {}^\forall M\in \alg{su}(2,2)\,. 
\end{align}
The generators $p_\mu$ used to parameterize $g\in SU(2,2)$ in 
\eqref{g-param} are defined as  
\begin{align}
p_\mu \equiv \frac{1}{2}\ga_\mu-m_{\mu 5}\,. 
\end{align}

\medskip 

To see the $\alg{so}(2,4)$ algebra explicitly, one can introduce the generators as follows:
\begin{align}
&\tilde m_{\mu\nu}=\frac{1}{4}\left[\ga_\mu,\ga_\nu\right]\,, \qquad 
\tilde m_{\mu5}=\frac{1}{4}\left[\ga_\mu,\ga_5\right] 
\qquad (\mu,\nu=0,1,2,3)\,, \nonumber \\ 
&\tilde m_{\mu, -1}=-\tilde m_{-1,\mu}=\frac{1}{2}\ga_\mu \,, \qquad 
\tilde m_{5, -1}=-\tilde m_{-1,5}=\frac{1}{2}\ga_5 \,.  
\end{align}
Indeed, these generators satisfy the commutation relations, 
\begin{align}
[\tilde m_{\hat{\mu}\hat{\nu}},\tilde m_{\hat{\rho}\hat{\sigma}}]
=\tilde \eta_{\hat{\rho}\hat{\nu}}\tilde m_{\hat{\mu}\hat{\sigma}}
-\tilde \eta_{\hat{\rho}\hat{\mu}}\tilde m_{\hat{\nu}\hat{\sigma}}
-\tilde \eta_{\hat{\sigma}\hat{\nu}}\tilde m_{\hat{\mu}\hat{\rho}}
+\tilde \eta_{\hat{\sigma}\hat{\mu}}\tilde m_{\hat{\nu}\hat{\rho}}\,. 
\end{align}
Here $\hat{\mu},\hat{\nu},\hat{\rho},\hat{\sigma}=-1,0,1,2,3,5$  
and $\tilde \eta_{\hat{\mu}\hat{\nu}}={\rm diag}(-1,-1,1,1,1,1)$\,.

\section{The T-duality rules} 
\label{app:Tdual} 

The rules of T-duality \cite{Buscher,T-duality} are summarized here.  
We basically follow Appendix C of \cite{T-duality}. 

\medskip 

The transformation rules between type IIB and type IIA supergravities 
are listed below. 
Note that the T-duality is performed for the $y$-direction 
and the other coordinates are denoted by $a,b,a_i~(i=1,\ldots)$\,. 
The fields of type IIB supergravity are the metric $g_{\mu\nu}$\,, 
NS-NS two-form $B_2$\,, dilaton $\Phi$\,, R-R gauge fields $C^{(2n)}$\,. 
The ones of type IIA supergravity are denoted with the tilde, 
the metric $\tilde{g}_{\mu\nu}$\,, NS-NS two-form $\tilde{B}_2$\,, dilaton $\tilde{\Phi}$\,, 
and R-R gauge fields $\tilde{C}^{(2n+1)}$\,.

\paragraph{From type IIB to type IIA}  
\begin{align}
\tilde g_{yy}&= \frac{1}{g_{yy}}\,, \qquad \tilde g_{ay}=\frac{B_{ay}}{g_{yy}}\,, 
\qquad \tilde g_{ab}=g_{ab}-\frac{g_{ya}g_{yb}-B_{ya}B_{yb}}{g_{yy}}\,, \el \\
\tilde B_{ay}&=\frac{g_{ay}}{g_{yy}}\,, \qquad \tilde B_{ab}=B_{ab}-\frac{g_{ya}B_{yb}-B_{ya}g_{yb}}{g_{yy}}\,,
\qquad 
\tilde \Phi=\Phi-\frac{1}{2}\ln g_{yy}\,, \nonumber \\ 
\tilde{C}^{(2n+1)}_{a_1\cdots a_{2n+1}} 
&= - C^{(2n+2)}_{a_1\cdots a_{2n+1}y} - (2n+1) B^{~}_{y[a_1}C^{(2n)}_{a_2\cdots a_{2n+1}]} 
+2n(2n+1)\frac{B^{~}_{y[a_1}g^{~}_{a_2 |y|} C^{(2n)}_{a_3\cdots a_{2n+1}]y}}{g_{yy}}\,, \nonumber \\ 
\tilde{C}^{(2n+1)}_{a_1\cdots a_{2n}y} &= C^{(2n)}_{a_1\cdots a_{2n}} 
+ 2n \frac{g^{~}_{y[a_1}C^{(2n)}_{a_2\cdots a_{2n}]y}}{g_{yy}}\,. 
\end{align}
where the anti-symmetrization for indices is defined as, for example,  
\[
A_{[a}B_{b]} \equiv \frac{1}{2}\left(A_a B_b - A_b B_a \right)\,. 
\]
The symbol $|y|$ inside the anti-symmetrization means that the indices other than the index $y$ are anti-symmetrized. 

\paragraph{From type IIA to type IIB}

\begin{align}
g_{yy}&= \frac{1}{\tilde{g}_{yy}}\,, \qquad g_{ay}=\frac{\tilde{B}_{ay}}{\tilde{g}_{yy}}\,, 
\qquad g_{ab} = \tilde{g}_{ab}-\frac{\tilde{g}_{ya}\tilde{g}_{yb}-\tilde{B}_{ya}\tilde{B}_{yb}}{\tilde{g}_{yy}}\,, \el \\
B_{ay}&=\frac{\tilde{g}_{ay}}{\tilde{g}_{yy}}\,, \qquad B_{ab} = \tilde{B}_{ab} - \frac{\tilde{g}_{ya}\tilde{B}_{yb} 
- \tilde{B}_{ya}\tilde{g}_{yb}}{\tilde{g}_{yy}}\,, \qquad 
\Phi = \tilde{\Phi}-\frac{1}{2}\ln \tilde{g}_{yy}\,, \nonumber \\ 
C^{(2n)}_{a_1\cdots a_{2n}} &= \tilde{C}^{(2n+1)}_{a_1\cdots a_{2n}y} - 2n \tilde{B}^{~}_{y[a_1}\tilde{C}^{(2n-1)}_{a_2\cdots a_{2n}]} 
+2n(2n-1)\frac{\tilde{B}^{~}_{y[a_1}\tilde{g}^{~}_{a_2|y|}\tilde{C}^{(2n-1)}_{a_3\cdots a_{2n}]y}}{\tilde{g}_{yy}}\,, \nonumber \\ 
C^{(2n)}_{a_1\cdots a_{2n-1}y} &= - \tilde{C}^{(2n-1)}_{a_1\cdots a_{2n-1}} 
- (2n-1)\frac{\tilde{g}^{~}_{y[a_1}\tilde{C}^{(2n-1)}_{a_2\cdots a_{2n-1}]y}}{\tilde{g}_{yy}}\,. 
\end{align}

\end{document}